\magnification=\magstep1
\hfuzz=6pt
\baselineskip=15pt
$ $
\vskip 1in

\centerline{\bf Quantum search without entanglement}

\bigskip

\centerline{Seth Lloyd}

\bigskip
\centerline{\it Department of Mechanical Engineering}

\centerline{\it MIT 3-160, Cambridge, Mass. 02139}

\centerline{\it slloyd@mit.edu}

\bigskip\noindent{\it Abstract:} Entanglement of quantum variables
is usually thought to be a prerequisite for obtaining quantum speed-ups
of information processing tasks such as searching databases.  This paper
presents methods for quantum search that give a speed-up over classical
methods, but that do not require entanglement.  These methods rely
instead on interference to provide a speed-up.  Search without entanglement
comes at a cost: although they outperform analogous classical devices,
the quantum devices that perform the search are not universal
quantum computers and require exponentially greater overhead than
a quantum computer that operates using entanglement.  Quantum search 
without entanglement is compared to classical search using waves.

\bigskip

Quantum computers exploit quantum coherence to perform computations
in ways that classical computers cannot.$^{1-5}$  Despite the 
considerable difficulties involved constructing quantum computers,$^{6-7}$ 
simple quantum logic devices have been built and prototype quantum 
computations have been performed.$^{8-16}$  Quantum computation is known 
to be able to solve some problems more rapidly than is possible 
classically.$^{17-24}$  Some problems, such as factoring and quantum 
simulation, can apparently be solved exponentially faster on a quantum 
computer than on a conventional digital computer.$^{19-21}$  Other 
problems, such as database search,$^{22-24}$ can be solved polynomially faster 
on a quantum computer.  The goal of this paper is to clarify 
what aspects of quantum mechanics are responsible for these speedups.  
In particular, it is often claimed that quantum speedups 
arise out of the quantum phenomenon known as entanglement.$^{25}$  
This paper shows that this claim, while accurate by and large, is 
incomplete: it is possible to obtain quantum speedups
using special-purpose devices that do not exhibit entanglement.
Grover's algorithm for database search, for example, searches 
a database with $n$ slots using only $O(\sqrt{n})$ queries,
while the best classical algorithm requires $O(n)$ queries.
Although Grover's algorithm as originally formulated induces
entanglement in the qubits of a quantum computer performing the
algorithm, this paper shows that it is possible to construct
quantum search devices that give the same $\sqrt{n}$ speedup over 
classical devices, but that do not require entanglement.  
These devices rely not on entanglement to obtain their quantum
speedup, but on interference.  Such devices are not general
purpose quantum computers: to perform quantum searches without
entanglement, they incur an exponentially greater overhead in
their incidental operations than a universal quantum computer
that operates using entanglement.  They nonetheless provide 
a speedup over classical devices.  Finally, the paper shows how
it is possible to construct classical search devices using waves
that provide a $\sqrt n$ speedup over the best classical
search device that uses particles.

Entanglement is a peculiarly quantum phenomenon that is responsible
for a variety of counterintuitive effects such as apparent quantum
nonlocality, quantum teleportation, etc.$^{26-27}$  A pure state
$|\psi\rangle$ for a quantum system composed of two or more 
subsystems is said to be entangled if it cannot be written in tensor 
product form:
$|\psi\rangle \neq |\psi_1\rangle \otimes |\psi_2\rangle \otimes \ldots $.
Note that entanglement is not a property of the state $|\psi\rangle$
on its own, but rather of the state and the way in which the system
is divided up into subsystems.  The claim that a quantum computation
requires entanglement relies on a division of the 
quantum computer into quantum bits or qubits.  

In Grover's algorithm
for database search, a single item located in one of $n$ slots in a 
database is located with only $O(\sqrt{n})$ queries of the database.$^{22}$
Here, a query is a question of the form, `Is the item in slot $x$?'
In classical database search $n-1$ queries are required in the 
worst case and $n/2$ queries are required on average.  
Grover's algorithm clearly gives a speedup over classical
search, and is normally taken to involve entanglement.
In Grover's original version of this algorithm, he took $n=2^r$
and performed the specified operations using quantum logic on
$r$ qubits.  For $r>2$ these operations entail putting the
qubits in an entangled state at some point in the operation.
However, as will now be shown, this entanglement is not an 
essential for obtaining a speedup over a classical device, 
but rather a byproduct of the mapping of the steps above 
onto qubits.  The following implementation allows one
to perform quantum search in a way that does not require entanglement, 
but that nonetheless does better than the best classical device.

Consider a box with $n$ slots through which a coin can be dropped.
In all but one of the slots, when the coin goes in heads it comes
out heads: the slot doesn't flip the coin.   In the remaining slot, 
when a coin goes in heads, it comes out tails: the slot flips the
coin.  The problem is to find which slot flips the coin.  One
way to find out is to take the box apart and look to see which
slot has a twist: but let's suppose that one is only allowed to
put coins in and see if they come out flipped or not.  In this case,
one has to put $n/2$ coins through on average and $n-1$ in the
worst case to locate the slot that flips the coin.  If there
is a meter on the box that charges a dollar for each coin
that goes through, searching a box with one hundred slots
costs \$50, on average.
This problem is clearly a version of the database search problem.
It differs from Grover's original formulation only in that
the Grover assumed the slots in his database to be labelled
by binary numbers, whereas in the `box' version no particular
labelling need be specified. 

Now look at a quantum mechanical version of this problem.  
Use a quantum particle such as a neutron as a quantum `coin': 
the box is constructed so that when a neutron with polarization
$\uparrow$ goes through all but one of 
the slots, it emerges with polarization $\uparrow$.
But when it goes through the remaining slot, it is flipped and emerges with
polarization $\downarrow$.  Equivalently, when it goes through
that slot with a polarization $\rightarrow~= (1/\sqrt 2)( \uparrow
- \downarrow)$, the neutron acquires a phase of $-1$.  For example, 
the slot that does the flipping could contain a magnetic field
along the $\rightarrow$ axis that flips the spin about that axis.
Let $|\ell\rangle$ be the state in which the neutron is in the
mode that goes throught the $\ell$'th slot.  Let $u$ be the label
of the slot that flips the neutron ($u$ for `unknown'). 
Using neutrons with polarization $\rightarrow$, the effect of the
box is to take the incoming state $|\ell\rangle \longrightarrow
(1-2\delta_{\ell u}) |\ell\rangle \equiv {\cal O} |\ell\rangle$.
${\cal O} = e^{-i\pi |u\rangle \langle u|}$ is the unitary operator 
that gives the effect of the box on the neutron.  (The letter ${\cal O}$
is used because the box effectively functions as what is called
in the field of computation an ${\cal O}$racle:
you ask it questions, it gives you answers, but its inner workings
are inscrutable.$^{18}$)

Now pose the question, `How many times must one put a neutron or
quantum coin through the quantum box to figure out which slot
flips the neutron?'  The answer is $O(\sqrt{n})$ times as the following
procedure shows.  We have $n$ translational modes of the neutron, 
one going through each
of the $n$ slots in the box.  Let ${\cal B}$ be the unitary operator
representing the action of a beam splitter that takes a neutron
from one mode and divides it equally amongst all the modes:
such a beam splitter can be constructed from $O(n)$ two-mode
beam splitters.  Let ${\cal B}^\dagger$ be the unitary operator 
corresponding to the `inverse' beam splitter that undoes the 
action of the first.  Finally, let ${\cal I} = - e^{-i\pi
|1\rangle \langle 1|}$ be the unitary operator that corresponds
to an inverter that gives every mode except for the first a phase of $-1$.
The inverter could also be constructed from a magnetic field.

The procedure for finding the unknown slot is as follows.  
Take a neutron in mode 1 and
put it in sequence through the beam splitter, then the box,
then the inverse beam splitter, then the inverter.  The
net effect is to apply the operator ${\cal I B^\dagger O B}$
to the initial state $|1\rangle$.  By comparison with
Grover's original algorithm, it is easily seen that
${\cal B}$ gives an action analogous to a Hadamard transformation
on the original state, ${\cal O}$ gives the same action
as Grover's `quantum phase oracle' that gives the effect
of querying the database, ${\cal B}^\dagger$ gives an
action analogous to an inverse Hadamard transformation, and ${\cal I}$
gives the same action as the operation Grover calls `inversion
about average.'  Now take the neutron and put it through the
beam splitter, box, inverse beam splitter, and inverter
again, and again, $O(\sqrt{n})$ times.  By the same calculation
as in Grover's original algorithm, the neutron is now with
high probability emerging from the $u$'th slot, and detection
of its position will reveal $u$.  The location of the slot
that flips the quantum coin has been determined by putting
a neutron through the box only $\sqrt{n}$ times.  If the box
has a meter that charges a dollar each time a neutron goes
through, searching a box with one hundred slots costs
only \$10.  The best classical strategy using coins, by comparison, 
costs on average \$50: the quantum version does better.  There is
no entanglement as there is only one neutron, 
and nothing for it to be entangled with.
  
The key to seeing that entanglement is not required for quantum search
is to note that there is nothing in Grover's description of his search
algorithm that requires the $n$-state system to be composed
of qubits.  The $n$ states could just as well be discrete states of 
a single quantum variable.  In such a `unary' representation
there is no entanglement as there is only one quantum variable.
There is nothing to be entangled with.  There is nothing in either
the classical or the quantum search problem that requires
that the problem be formulated in a `binary' representation in which
the slots in the database are labeled by binary numbers.  Indeed,
the unary representation of the search problem is more
`natural' in that the slots are labeled by natural numbers without
requiring that a particular base (2 or 10, e.g.) be specified.  
Even if one demands that a base be specified, then as long as
the base is greater than $n$, no entanglement is required.
In fact, the point that entanglement is not
required in few-qubit quantum algorithms has been noted before,$^{28}$
but with the misleading conclusion that the algorithms are not
quantum-mechanical because entanglement is not present.
Clearly, the unary representation still gives a $\sqrt{n}$ 
speedup over the classical search problem in the sense that in the
quantum version the quantum `coin' need only be passed
through the box $\sqrt{n}$ times.

The use of a unary representation does not come without cost,
however.  The conventional binary version of Grover's algorithm
uses $O(\log_2 n)$ qubits and requires $O(\log_2 n)$ operations
to perform each inversion about average and to determine the
final result.  The unary version of Grover's algorithm, in
contrast, although it requires only $O(1)$ operation to
perform the inversion about average (all that is required
is a single phase delay on the first mode) requires $O(n)$ 
two-mode beam splitters to manipulate the neutron and $O(n)$ detectors 
to read out the result.  Although both devices give a $\sqrt n$ 
speedup over the analogous classical device in the sense that
they have to consult the `oracle' fewer times, the 
unary version requires exponentially more resources than the qubit
version to perform the incidental operations.  Like
unary optical simulators of quantum logic,$^{29}$ such devices 
are emphatically not universal quantum computers.  The number of 
resources required to simulate an $N$ qubit quantum computation
using such a unary representation goes as $2^N$.  Accordingly,
unary devices cannot provide an exponential speedup over classical
devices: the best they can do is reduce the number of times
that they consult the database or `oracle.'  Problems that
do not involve consulting an oracle cannot be sped up by such devices.  
To map the operation of Shor's algorithm to a unary device, or
to simulate a quantum system on such a device would require 
exponential resources.  

Before turning to classical search using waves, a further discussion
of entanglement is in order.
As noted above, entanglement is not a property of the state of a 
system on its own, but rather of the state and the way in which 
one divides the system up into subsystems.  By changing the
way one divides up the system, it is always possible to represent
an unentangled state as an entangled state and {\it vice versa}.  For example,
if one describes the single neutron interferometric database search
method in a `second quantized' picture, in which the state 
$|0\rangle_1 \ldots |1\rangle_\ell \ldots |0\rangle_n$ represents 
a state in which the neutron is in the $\ell$'th mode, then the
initial state $(1/\sqrt n) (|1\rangle_1 \ldots |0\rangle_n +
\ldots + |0\rangle_1 \ldots |1\rangle_n)$ exhibits entanglement
between the $n$ modes.  We now present two further unary versions of 
quantum in which such a second quantized picture is less applicable.  

First, the $n$ states could be different 
energy levels of a single atom.  In this case, the action of
${\cal B}$ above could be accomplished by a shaped,
broadband pulse that takes the atom from the state
$|1\rangle$ to an equal superposition of the first $n$ energy
levels; the box could effect the phase inversion ${\cal O}$
of the unknown state $|u\rangle$ by driving a $2\pi$ pulse
between $|u\rangle$ and the ground state $|0\rangle$ (recall that we're
not allowed to look inside the box and determine $u$ by detecting
this pulse); the action of $B^\dagger$
could be accomplished by a shaped, broadband pulse that
inverts ${\cal B}$; and the inversion about average ${\cal I}$
could be accomplished by driving a $2\pi$ pulse between
$|1\rangle$ and $|0\rangle$ as for ${\cal O}$ followed 
by a broadband $2\pi$ pulse between all the states and
the ground state.  After $O(\sqrt{n})$ iterations of
the operations ${\cal I B^\dagger O B}$, the system
is in the state $|u\rangle$ and a measurement of its energy
will reveal the position in the database.  This measurement could
be performed, for example, by interchanging each state $|x\rangle$ with
the ground state in turn, and by driving a cycling transition
that induces fluorescence if and only if the system is in the
ground state.  Although such a measurement requires up to $n$
steps, it does not require any further passages through the box.

A second example of database search without entanglement is
given by the Farhi-Gutmann continuous version of Grover's
algorithm.$^{24}$  Here, the system could be a spin with $n$ states,
and the database is given by a box that
applies a Hamiltonian $|u\rangle \langle u|$.  You are allowed
to prepare the spin any desired state, and to add your
own preferred Hamiltonian to the unknown Hamiltonian applied
by the box.  Farhi and Gutmann show that if you start the spin
in an {\it arbitrary} state $|\psi\rangle$ and add to the
database Hamiltonian the Hamiltonian $|\psi\rangle \langle \psi|$
then after a period of time proportional to $\sqrt{n}$ the
spin has rotated to the state $|u\rangle$ with high probability. 

Once it is clear that a representation in terms of qubits is
not required for quantum search, many implementations are possible.
In fact, as will now be seen, a {\it classical} implementation 
phrased in terms of waves can still give a $\sqrt n$ speedup over the
classical search problem phrased in terms of coins or particles.
The set of search methods has now come full circle: it was by 
considering interference via classical waves emitted by an array 
of antennae that Grover arrived at his quantum algorithm in the 
first place.$^{30}$

Return now to the neutron interferometer picture of quantum
search described above.  What in this picture is quantum-mechanical?
There are in fact only two points in which the quantum nature
of the neutron appears.  The first is in the billing procedure:
the box charges on a per quantum basis.  The second is the
final click of the detector at the slot from which the neutron
emerges.  These are the only points at which the neutron is required
to behave like a particle.  At all other points in the search process, 
it is the wave aspect of the neutron that comes into play: it
is the interference between the waves in the interferometer
that lies behind the $\sqrt n$ speedup.   

This dominance of the wave aspect of quantum mechanics suggests
the following purely classical wave method for search.  Instead
of quantum matter waves, use classical waves such as light or
sound.  At bottom, of course, such waves are composed of photons and
phonons.  But it is possible to reformulate the search problem
in such a way that the particle aspect of the waves is unimportant.
Let the unknown slot in the box flip the polarization of the
waves, and suppose now that the box charges on the
basis of the integrated intensity of the waves that pass
through it rather than on a per particle basis.  We are provided
with detectors with a finite signal to noise ratio.  How now
does the cost of determining the unknown slot scale with
the number of slots $n$?    

One way to search the box is simply to shine waves through
all the slots at once and to determine which slot flips
the polarization of the transmitted wave.  Because of
the finite signal to noise ratio of the detectors, the
cost of this method is proportional to $n$.  A second method
is to recycle the waves through an interferometer constructed
in exact analog to the neutron interferometer described above 
to give positive interference at the output of the unknown slot.
Just as in the quantum case, the cost of this method is proportional
to $\sqrt n$.  So a purely classical wave search device can also
find the unknown slot with an integrated intensity proportional
to $\sqrt n$.  Of course, if one tries to minimize costs
by decreasing the intensity of the recycled waves and increasing 
the sensitivity of the detectors, one's data will eventually
arrive in the form of individual `clicks': the quantum
nature of the wave will reassert itself. 

To summarize: 

\noindent$\bullet$ A classical digital computer that searches a database
with $n$ slots requires $O(\log_2 n)$ resources and has to
look at the database $O(n)$ times.  

\noindent$\bullet$ A quantum digital computer
that searches a database with $n$ slots requires $O(\log_2 n)$ 
resources and has to look at the database $O(\sqrt n)$ times.  

\noindent$\bullet$ A classical device that determines which of 
$n$ slots in a box flips a discrete object such as a coin requires  
$O(n)$ resources and has to pass the coin through $O(n)$ times.

\noindent$\bullet$ A quantum device that determines which of $n$ slots 
in a box flips a discrete object such as a particle requires  
$O(n)$ resources and has to pass the particle through $O(\sqrt n)$ times.  

\noindent$\bullet$ A classical wave device that determines which of $n$ slots 
in a box flips the polarization of a wave requires  
$O(n)$ resources and has to send the wave through $O(\sqrt n)$ times.

Special purpose quantum search devices can give a 
speedup over classical search devices without using entanglement.  
A quantum device that probes a system by sending discrete objects
such as particles through it can acquire information about 
unknown features of the system more rapidly than analogous classical 
devices that probe a system by sending discrete objects through it.  
The $\sqrt n$ speedup obtained by the quantum devices arises 
out of the wave nature of the particles sent through. 
Classical devices that rely on waves and
interference can also give a $\sqrt n$ speedup over classical
devices that probe a system using particles alone. 

\vfill
\noindent{\it Acknowledgements:} The author would like to thank
H.J. Kimble for pointing out the essential distinction between
quantum search using particles and classical search using waves.
T. Weinacht suggested the possibility of performing quantum search 
in atoms.  S. Braunstein,  S. van Enk, R. Jozsa, and M. Knill 
contributed by helpful discussions.  This work was supported by
DARPA under the QUIC initiative. 
\eject

\centerline{\bf References}

\item{1.} P. Benioff, {\it J. Stat. Phys.} {\bf 22}, 563 (1980);  {\it
Phys. Rev. Lett.} {\bf 48}, 1581 (1982);  {\it J. Stat. Phys.} {\bf
29}, 515 (1982);  {\it Ann. N.Y. Acad. Sci.} {\bf 480}, 475 (1986).

\item{2.} R.P. Feynman, {\it Opt. News} {\bf 11}, 11 (1985); {\it
Found. Phys.} {\bf 16}, 507 (1986); {\it Int. J. Theor. Phys.} {\bf
21}, 467 (1982).

\item{3.} D. Deutsch, {\it Proc. R. Soc. London Ser. A} {\bf 400},
97 (1985).

\item{4.} D.P. DiVincenzo, {\it Science} {\bf 270}, 255 (1995).

\item{5.} S. Lloyd, {\it Sci. Am.} {\bf 273} (4), 140 (1995).

\item{6.} R. Landauer, {\it Int. J. Theor. Phys.} {\bf 21}, 283
(1982); {\it Found. Phys.} {\bf 16}, 551 (1986);
{\it Nature} {\bf 335}, 779 (1988);
{\it Nanostructure Physics and Fabrication}. M.A. Reed and
W.P. Kirk, eds. (Academic Press, Boston, 1989), pp. 17-29;
{\it Physics Today} {\bf 42}, 119 (October 1989); {\it Proc. 3rd
Int. Symp. Foundations of Quantum Mechanics, Tokyo}, 407 (1989);
{\it
Physica A} {\bf 168}, 75 (1990); {\it Physics Today}, 23 (May 1991);
{\it Proc. Workshop on Physics of Computation II}, D. Matzke ed., 1 ({\it
IEEE} Press, 1992).

\item{7.} S. Lloyd, {\it Science} {\bf 261}, 1569 (1993); {\it
Science} {\bf 263}, 695 (1994).

\item{8.} Q.A. Turchette, C.J. Hood, W. Lange, H. Mabuchi, H.J.
Kimble, {\it Phys. Rev. Lett.} {\bf 75}, 4710, (1995).

\item{9.} C. Monroe, D.M. Meekhof, B.E. King, W.M. Itano, D.J.
Wineland, {\it Phys. Rev. Lett.} {\bf 75}, 4714, (1995).

\item{10.} D.G. Cory, A.F. Fahmy, T.F. Havel, in {\it PhysComp96}, 
Proceedings of the
Fourth Workshop on Physics and Computation, T. Toffoli, M. Biafore,
J. Le\~ao, eds., New England Complex Systems Institute, 1996,
pp. 87-91; {\it Proc. Nat. Acad. Sci.} {\bf 94}, 1634 (1997);
{\it Physica D} {\bf 120}, 82 (1998).

\item{11.} N.A. Gershenfeld and I.L. Chuang, {\it Science} 
{\bf 275}, 350-356 (1997).

\item{12.} I.L. Chuang, N. Gershenfeld, M.G. Kubinec, D.W. Leung,
{\it Proc. R. Soc. Lond. A} {\bf 454}, 447 (1998).

\item{13.} I.L. Chuang, L.M.K. Vandersypen, X. Zhou, D.W. Leung, S. Lloyd,
{\it Nature} {\bf 393}, 143 (1998).

\item{14.} I.L. Chuang, N. Gershenfeld, M.G. Kubinec, {\it Phys. Rev. Lett.}
{\bf 80}, 3408 (1998).

\item{15.} J.A. Jones, M. Mosca, {\it J. Chem. Phys.} {\bf 109}, 
1648 (1998).

\item{16.} J.A. Jones, M. Mosca, R.H. Hansen, {\it Nature} {\bf 393} 
344 (1998).

\item{17.}  D. Deutsch and R. Jozsa, {\it Proc. R. Soc. London A} 
{\bf 439}, 553 (1992) 

\item{18.} E. Bernstein and U. Vazirani, 
in {\it Proceedings of the 25th Annual ACM Symposium
on the Theory of Computing}, ACM Press, New York, 1993, pp. 11-20.

\item{19.} P. Shor, in {\it Proceedings of the 35th Annual Symposium on
Foundations of Computer Science}, S. Goldwasser, ed., ({\it IEEE} Computer
Society, Los Alamitos, CA, 1994), p. 124.

\item{20.} D.R. Simon,
in {\it Proceedings of the 35th Annual Symposium on Foundations
of Computer Science}, S. Goldwasser, Ed., IEEE Computer
Society, Los Alamitos, CA, 1994, pp. 116-123.

\item{21.} S. Lloyd, {\it Science}, Vol. {\bf 273}, pp. 1073-1078 (1996).
See also S. Wiesner, eprint quant-ph (1996) and C. Zalka, eprint (1996).

\item{22.} L.K. Grover, {\it Phys. Rev. Lett.},
Vol. {\bf 79}, pp. 325-328 (1997);
in {\it Proceedings of the 28th Annual ACM Symposium
on the Theory of Computing}, ACM Press, New York, 1996, pp. 212-218.

\item{23.} B.M. Terhal, J.A. Smolin, {\it Phys. Rev. A}
{\bf 58}, 1822 (1998).

\item{24.} E. Farhi, S. Gutmann, {\it Phys. Rev. A} {\bf 57}, 2403 (1998).

\item{25.} R. Jozsa, in {\it Geometeric Issues and the Foundations
of Science,} eds. S. Huggett, L. Mason, K.P. Tod, S.T. Tsou,
and N.M.J. Woodhouse, Oxford University Press (1997); R. Jozsa and A. Ekert,
quant-ph 9803072 (1999), to appear in {\it Phil. Trans. Roy. Soc.}.

\item{26.} For a review of entanglement, its properties and uses,
see A. Peres, {\it Quantum Theory: Concepts and Methods},
(Kluwer, Dordrecht 1993). 

\item{27.} C.H. Bennett, {\it Physics Today} {\bf 48}, 24-30
(1995).

\item{28.} D. Collins, K.W. Kim., W.C. Holton, {\it Physical Review A} 
{\bf 58}, pp. R1633-R1636 (1998).

\item{29.} N.J. Cerf, C. Adami, and P.G. Kwiat, {\it Phys. Rev. A}
{\bf 57}, 1477 (1998).

\item{30.} L. Grover, private communication.

\vfill\eject\end